\documentclass[12pt]{emulateapj} 
\usepackage{graphicx} \usepackage{multirow} \usepackage{color}
\usepackage{soul} 

\usepackage{ot1patch} \usepackage{color} \usepackage[dvipsnames]{xcolor}
\usepackage{tikz}

\advance \voffset by 1.5cm\relax

\newcommand{\msun}{\ensuremath{\,\mathrm{M}_\odot}}
\newcommand{\yr}{\ensuremath{\,\mathrm{yr}}}
\newcommand{\myr}{\ensuremath{\,\mathrm{Myr}}}

\newcommand{\rsun}{\ensuremath{\,\mathrm{R}_\odot}}
\newcommand{\startrack}{{\tt StarTrack }} 
\newcommand{\s}{\ensuremath{\S\,}}
\newcommand{\msy}{\ensuremath{\msun\mathrm{\; yr}^{-1}}}
\newcommand{\ergs}{\ensuremath{\,\mathrm{erg}\,\mathrm{s}^{-1}}}
\newcommand{\lx}{\ensuremath{L_{\rm X}}} 
 
\newcommand{\zsun}{\ensuremath{\mathrm{Z}_\odot}}
\newcommand{\lsun}{\ensuremath{\mathrm{L}_\odot}} 
\newcommand{\mpc}{\ensuremath{\,\mathrm{Mpc}}}

\newcommand{\mdotyr}{\ensuremath{\mdot\,\mathrm{yr}^{-1}}}
\newcommand{\facc}{\ensuremath{f_{\rm acc}}}
\newcommand{\fminus}{\ensuremath{f^-}}
\newcommand{\fplus}{\ensuremath{f^+}}
\newcommand{\macc}{\ensuremath{\mdot_{\rm acc}}}
\newcommand{\mrlof}{\ensuremath{\mdot_{\rm RLOF}}}
\def\mdot{\dot M} 

\newcommand{\new}[1]{\textbf{#1}}

\defcitealias{Sadowski15}{S15}
\defcitealias{Ohsuga07}{O07}

\begin{document}

\title{Nature of the Extreme Ultraluminous X-ray Sources}

\author{Grzegorz Wiktorowicz\altaffilmark{1},  Ma\l{}gorzata
Sobolewska\altaffilmark{2}, Aleksander S\k{a}dowski\altaffilmark{3}, and Krzysztof
Belczynski\altaffilmark{1, 4}} 

 \affil{ $^{1}$ Astronomical Observatory, University of Warsaw, Al.  Ujazdowskie
 4, 00-478 Warsaw, Poland (gwiktoro@astrouw.edu.pl)\\ $^{2}$ Nicolaus
 Copernicus Astronomical Center, Bartycka 18, 00-716 Warsaw, Poland\\ $^{3}$
 MIT Kavli Institute for Astrophysics and Space Research 77 Massachusetts Ave,
 Cambridge, MA 02139, USA\\ $^{4}$ Warsaw Virgo Group }
 
\begin{abstract} 
    In this proof-of-concept study we demonstrate that in a binary system mass
can be transferred toward an accreting compact object at extremely
high rate. If the transferred mass is efficiently converted to X-ray
luminosity (with disregard of the classical Eddington limit) or if the 
X-rays are focused into a narrow beam then binaries can form extreme ULX 
sources with the X-ray luminosity of $\lx \gtrsim 10^{42}\ergs$.
For example, Lasota \& King argued that the brightest known ULX (HLX-1) is a
regular binary system with a rather low-mass compact object (a stellar-origin
black hole or a neutron star).
The predicted formation efficiencies and lifetimes of binaries with the very 
high mass transfer rates are large enough to explain {\em all} observed systems 
with extreme X-ray luminosities. These systems are not only limited to binaries 
with stellar-origin black hole accretors. Noteworthy, we have also identified such 
objects with neutron stars. Typically, a $10\msun$ black hole is fed by a massive 
($\sim10\msun$) Hertzsprung gap donor with Roche lobe overflow rate of 
$\sim 10^{-3}\msy$ $(\approx 2600 \dot M_{\rm Edd}$). For neutron star systems the 
typical donors are evolved low-mass ($\sim2\msun$) helium stars with Roche lobe 
overflow rate of $\sim 10^{-2}\msy$.
Our study does not prove that any particular extreme ULX is a regular binary
system, but it demonstrates that {\em any} ULX, including the most luminous
ones, may potentially be a short-lived phase in the life of a binary star.
\end{abstract}


\section{Introduction}

Our Universe is populated with black holes (BH) and neutron stars (NS) in
various binary configurations. In our Galaxy, many such binaries show
significant X-ray activity suggestive of a mass transfer and accretion onto the
compact star. X-ray luminosities of only two Galactic X-ray binaries (XRBs) exceed
$\sim$10$^{39} \ergs$ (approximately the Eddington limit for a 10\,${\rm
M}_{\odot}$ BH) in GRS 1915+105 \citep{Fender04} and possibly in SS 433 if
the system were observed along the jet axis \citep{Fabrika06}.

However, a large population of extra-galactic point-like X-ray sources with
luminosities in excess of $\sim$10$^{39}\,\ergs$ has been identified
\citep[e.g.,][]{Fabbiano89, Liu11, Walton11}.  These so-called Ultraluminous
X-ray sources (ULX) are off-nuclear and therefore accretion onto a supermassive
BH ($M>10^5\,{\rm M}_{\odot}$) can be excluded as the source of their
luminosity. Instead, the two most popular scenarios to explain their nature
include binary systems hosting {\em (i)} a stellar mass BH accreting at a
super-Eddington rate, or {\em (ii)} an intermediate mass BH (IMBH) and
sub-Eddington accretion \citep{Colbert99}. In the latter case, formation of BHs
heavier than $\sim100\,{\rm M}_{\odot}$ presents a problem for current models
of stellar evolution. Although, it has been suggested that such IMBHs may be
formed in dense globular clusters \citep{Miller02} or even as a result of
stellar evolution of very massive stars \citep[$\sim
200-300\msun$;][]{Crowther10, Yusof13}. 
 
The super-Eddington BH accretion invoked in the stellar origin scenario remains
still a relatively poorly understood regime. However, several theoretical
mechanisms able to breach the Eddington limit have been proposed, e.g., beaming
and/or hyper-accretion allowed by non-uniform escape of photons from accretion
flow (''photonic bubbles'') \citep{King01, Begelman06, Poutanen07, King08}, as
well as a contribution of rotation powered pulsars and pulsar wind nebulae to
the ULX population \citep{Medvedev13}.

Robust observational constrains on the mass of the accretor in some ULXs
\citep[e.g.,][]{Motch14} indicate that indeed super-Eddington accretion
onto stellar-origin BHs is realised in nature. Moreover, \citet{Bachetti14}
investigating the M82 X-2 source have demonstrated that the
super-Eddington accretion is also possible in XRBs hosting a NS.  Additionally,
some sources transit relatively fast between the super- and sub-Eddington
regimes on timescales as short as a few days to a week
\citep[e.g.,][]{Walton13, Bachetti13}. Such short timescales are in
contradiction with IMBH accretors \citep{Lasota15b}.

Nevertheless, it has been speculated that the brightest ULXs, with
luminosities $>10^{41}\,\ergs$ may be candidate IMBHs \citep[e.g.,][]{Sutton12}. 
Recently, compelling evidence, based on quasi-periodic oscillations (QPO), was 
presented in support of a $\sim400\msun$ BH in M82 X-1 \citep{Pasham14}. However, 
it was also suggested that these QPOs may be harmonics of pulsar rotation period 
\citep{Kluzniak15}.
Moreover, \citet{Sutton15} demonstrated that the IMBH candidate in IC 4320 is
actually a background AGN. To date HLX-1 with $\lx =1.1\times10^{42}\,\ergs$
\citep{Farrell11} is the brightest known ULX \citep[for a discussion of the
brightest ULXs see][]{Servillat11}.

We approach the ULX issue from the standpoint of one particular evolutionary
model for binary evolution. We consider only the far end of the ULX luminosity
space, $\lx \gtrsim 10^{42}\ergs$, and we refer to the sources potentially able
to reach these luminosities as extreme ULXs (EULXs). We explore the possibility
that EULXs are binary systems with Roche lobe overflow mass transfer rates that 
highly exceed classical Eddington limit. For the purpose of this proof-of-concept 
study, we assume that the transferred mass is efficiently accreted onto a
compact object and converted to X-ray luminosity in the full range of possible 
mass accretion rates. This is in contrast with the generally accepted view that 
the conversion efficiency decreases with increasing mass accretion rate 
\citep[e.g.,][]{Poutanen07}. However, if mass is lost during accretion process 
and even if conversion into X-ray luminosity is not fully efficient, a geometrical 
beaming can provide large X-ray luminosities for sources considered in our study.

\section{Model}\label{sec:model}

We employ the binary population synthesis code, \startrack
\citep{Belczynski08}, with updates as discussed in \citet{Dominik14}. We
employed following initial conditions: \citet{Kroupa93} broken power law for initial mass function
(primary mass between $6$ -- $150\msun$), flat mass ratio distribution
\citep[secondary mass between $0.08$ -- $150\msun$]{Kobulnicky06}, flat in
logarithm distribution of separations \citep[$f(a)\sim 1/a$]{Abt83} and thermal distribution of
eccentricities \citep[$f(e)\sim e$]{Duquennoy91}.

We use
the {\tt BOINC} platform\footnote{http://boinc.berkeley.edu/} for volunteer
computing  in our program {\tt ``Universe@home''} (http://universeathome.pl) to
obtain a large number of X-ray binaries ($N=10^9$ of massive binary systems
were evolved by volunteers).  The X-ray binary is defined as a system hosting a
donor star transferring mass via Roche lobe overflow (RLOF) to a compact object
companion (NS or BH). For any given system our evolutionary models provide the
donor RLOF mass transfer rate, $\mdot_{\rm RLOF}$.  The accretion rate
onto the compact object, $\mdot_{\rm acc}$, is estimated in three
different ways from the mass transfer rate (see below). 
 
We search for evolutionary channels that allow for the formation of binaries
with highest possible RLOF mass transfer rates onto stellar-origin BHs and
NSs. Conservatively, we model only solar metallicity
\citep[$\zsun=0.02$;][]{Villante14} and we allow initial mass function (IMF) to
extend only to $150\msun$.  Note that lower metallicity and higher mass stars
may form massive ($\gtrsim 100\msun$) black holes in binary systems
\citep{Belczynski14}.

\begin{figure}
    \centering
    \includegraphics[scale=1.8]{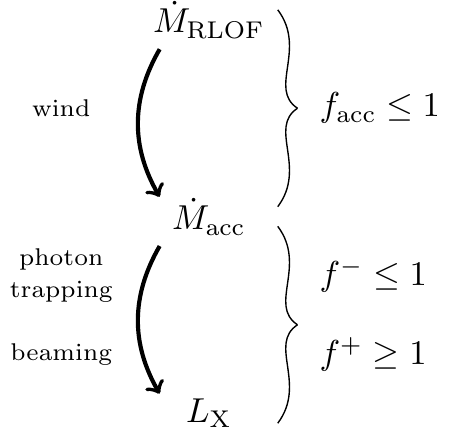}
    \caption{Most important processes behind conversion of RLOF mass
    transfer rate \mrlof\ into X-ray luminosity. Our parametrisation of these 
    processes is discussed in detail in \S\ref{sec:model} and used in Eq.~\ref{eq:l_x}.}
    \label{fig:graph}
\end{figure}

To estimate RLOF mass transfer rate \mrlof\ we evaluate donor properties
and various angular momentum loss mechanisms in a given binary as described in
\citet{Belczynski08}. It is highly uncertain how RLOF mass transfer rate is to
be  converted to X-ray luminosity, and we discuss this issue below. 

The gravitational energy of the RLOFing material becomes converted into
radiation in an accretion disk formed around the compact object \citep[see
\citet{Lasota15b} for a recent review on accretion disk physics]{Shakura73}. 
There are a number of effects that have a pivotal influence on the \mrlof\
to \lx\ conversion process in high mass-accretion rate disks (see Fig.
\ref{fig:graph}). 

The first conundrum to consider is the role of winds launched from the
disk surface.  Such winds have been ubiquitously detected in high mass
accretion rate disk-dominated states of XRBs \citep[e.g.,][]{Ponti12,Ponti15}.
Winds are able to remove a substantial fraction of matter that is being
transferred through the disk toward the compact object. At the same time, winds
carry away the angular momentum and therefore influence the orbital evolution.
We introduce the quantity \facc\ to describe the fraction of \mrlof\ that is
not affected by the disk winds (\macc). 

Secondly, not all of the photons produced in the vicinity of the accretor
get emitted from the disk. Some of them are dragged by the inflowing matter and
fall onto the compact object. As a result of this "photon trapping" effect
(e.g., \citeauthor{Ohsuga07} \citeyear{Ohsuga07}, hereafter
\citetalias{Ohsuga07}; \citeauthor{Abramowicz08} \citeyear{Abramowicz08};
\citeauthor{Narayan12} \citeyear{Narayan12}; \citeauthor{Sadowski15}
\citeyear{Sadowski15}, hereafter \citetalias{Sadowski15}), the observed X-ray
luminosity is reduced. To include this and other currently unknown processes
that may lower \lx\ we introduced the \fminus\ parameter.

Finally, we took into account the processes that may increase the
observed \lx\ due to non-isotropic emission. We utilise the \fplus\ factor
which in addition to the beaming \citep{King01} may also include a contribution
due to disk geometry, black hole spin, column accretion in magnetised NS, etc.
The beamed emission will always exceed the isotropic one ($\fplus>1$; as an
example, if the emission goes into a cone of opening angle $1^\circ$,
$10^\circ$ or $100^\circ$, it will correspond to \fplus\ $\sim2.6\times10^{4}$,
$\sim260$, and $\sim2.8$, respectively). However, a beamed source will be
visible only from the directions enclosed by the cone, and thus its detection
probability will be lower than that of an isotropic emitter. 

The XRB X-ray luminosity is calculated as\\
\begin{equation}\label{eq:l_x} 
    \lx = \fminus\fplus \frac{\epsilon G M_{\rm acc}\; \facc \mrlof}{R_{\rm
    acc}} = \eta\mrlof c^2, 
\end{equation} 
where the radius of the accretor, $R_{\rm acc}$, is 10 km for a NS and 3
Schwarzschild radii for a BH, $\epsilon$ gives a conversion efficiency of
gravitational binding energy to radiation associated with accretion onto a NS
(surface accretion $\epsilon=1.0$) and onto a BH (disk accretion
$\epsilon=0.5$), $\eta=\fminus\fplus\facc\eta_0$ is the total efficiency of the
accretion flow, and $\eta_0$ is the radiative efficiency of a standard thin
disk. \footnote{In our case $\eta_{0,BH}=1/12$ and $\eta_{0,NS}\approx0.2$
\citep[e.g.,][]{Shakura73}.}

In our simulations we define a potential EULX source as the one with 
$\lx >10^{42}$ erg s$^{-1}$.

\subsection{The reference model}\label{sec:reference}

In our reference model we assume: no winds ($\facc=1$), no photon trapping
($\fminus=1$), no beaming ($\fplus=1$), and always efficient mass accretion
rate-into-luminosity conversion ($\eta_0$ as in the standard thin disk).  We
adopt this condition even for very high RLOF mass transfer rates. This is
quite arbitrary, however, our goal was to estimate the highest luminosities
potentially reachable by an X-ray binary system without invoking strong
beaming. On the other hand, $\eta=\eta_0$ may also correspond to a situation
when a significant part of the \mrlof\ is lost from the system ($\facc<1$) or
\lx\ is lowered due to photon trapping ($\fminus<1$) but the beaming
compensates these effects ($\fplus=1/\facc\fminus$). Numerical simulations
\citepalias[e.g.,][]{Sadowski15} suggest that \facc\ is indeed small. However,
\fplus may reach $10^3$--$10^4$ for the cone opening angles of the order of a
few degrees.

Therefore, with our reference model we obtain the maximum potential X-ray
luminosity of a given binary system if the non-standard accretion disk effects
are negligible or compensate each other.

\subsection{The Ohsuga model}\label{sec:model_ohsuga}

Next, we have considered a BH accretion case with $\dot{M}_{\rm acc} <
\dot{M}_{\rm RLOF}$ ($\facc<1$), and $\dot{M}_{\rm acc}$ constrained following
the results of \citetalias{Ohsuga07} who performed global, axisymmetric
simulations of super-critical disks with the $\alpha P$ viscosity, including
effects of outflowing winds and photon trapping. We use the parametrisation of
the \citetalias{Ohsuga07} results (\macc\ as a function of \mrlof) derived in
\citet{Belczynski08a},

\begin{equation}\label{eq:macc_ohsuga} 
\log\left( \frac{\macc}{\mdot_{\rm crit}}\right) =
0.934\log\left(\frac{|\mdot_{\rm RLOF}|}{\mdot_{\rm crit}} \right) - 0.380.
\end{equation}
where $\mdot_{\rm crit}=2.6\times10^{-8}({\rm M}_{\rm BH}/10 \msun) \msy$
is the critical mass accretion rate. To obtain X-ray luminosities we
utilised the Eq.~\ref{eq:l_x} using \macc\ as provided by
Eq.~\ref{eq:macc_ohsuga} and substituting $\fminus\fplus\facc\mrlof=\macc$.
For a $10\msun$ BH accretor we obtain $2.8$ times lower luminosity if
$\mrlof=10^{-7}\mdotyr$ and $6.7$ times lower if $\mdot_{\rm RLOF}=0.1
\mdotyr$. For illustration, if the collimation angle of the outflow is
$\sim20^\circ$, as in NS simulations of \citet{Ohsuga07b}, the corresponding
$\fplus$ will be equal $66$. The accretion onto NSs is limited to the
classical Eddington limit in this model \citep{Ohsuga07b}. $\lx$ is calculated
as in Eq.~\ref{eq:l_x}. We note that we have extrapolated results of the original calculations of \citet{Ohsuga07} to the range of mass transfer rates we have in our simulations.

\subsection{The S\k{a}dowski model}

Finally, we have experimented with the constraints on the $\dot{M}_{\rm
acc}$ and $L_{\rm x}$ following the recent results of \citetalias{Sadowski15}.
The accretion rate at the BH horizon was found to be only a fraction of the
RLOF rate, 

\begin{equation} \label{eq:mdot_as} 
\macc = \mrlof\frac{R_{\rm in}}{R_{\rm out}},
\end{equation}
where $R_{\rm out}$ and $R_{\rm in}=20R_{\rm g}$ \citep{sadowski+koral2}
are the outer and inner radii of the wind emitting region, respectively. It is
hard to estimate the location of the outer edge, so we assumed a constant
fraction $R_{\rm wind} / R_{\rm out} = 0.01$ (the wind probably is not emitted
out to the edge of the disk), which corresponds to $\facc=0.01$. In this
model we utilised a different formula for X-ray luminosity which already
accounts for the effects of photon trapping and beaming,

\begin{equation} \label{eq:l_as}
\lx=4\times10^{38}\times e^{-\frac{\theta}{0.2}}\frac{\dot{M}_{\rm
acc}}{\dot{M}_{\rm Edd}} \frac{M_{\rm BH}}{M_\odot}\ergs, 
\end{equation}
where $\theta$ is the viewing angle, $\dot{M}_{\rm
Edd}=2.44\times10^{18}\frac{M}{M_{\odot}}{\rm\, g\;s^{-1}}$ is the Eddington
accretion rate. In our simulations we incorporated $\theta$ in the range $0^\circ$ --
$30^\circ$ (the opening angle is $60^\circ$). This will correspond to
$\fplus\approx8$ in our reference model. In \citet{Sadowski15} model we get different
luminosities for different viewing angles, which was included in our
simulations to calculate probabilities of observing a particular system as the
EULX. Even though the \citetalias{Sadowski15} model was constructed for systems
with BHs, we assumed that the same prescription is valid for the NSs accretors.
We note that the results of \citetalias{Ohsuga07} can be put in the framework
of this model by taking $R_{\rm out}\approx 50 R_{\rm g}$, which is
approximately the effective circularization radius (and the outer edge of the
disk) of the gas injected into their simulation box. We note that the Eq.~\ref{eq:l_as} was obtained by \citet{Sadowski15} for supermassive BH, but we extrapolated it to the stellar mass ones.

\section{Results} \label{sec:results} 

Under the least restricting assumption (no limit on accretion rate,
$\eta=\eta_0$) our EULX rate/number estimates are to be considered upper
limits.  We find that $1$ per $44$ billion binaries could potentially be 
an EULX with a BH accretor, and $1$ per $44$ billion binaries could
potentially be an EULX with a NS accretor. This estimate employs a canonical 
IMF \citep{Kroupa03}.

\begin{figure}
    \centering
    \includegraphics[scale=0.5]{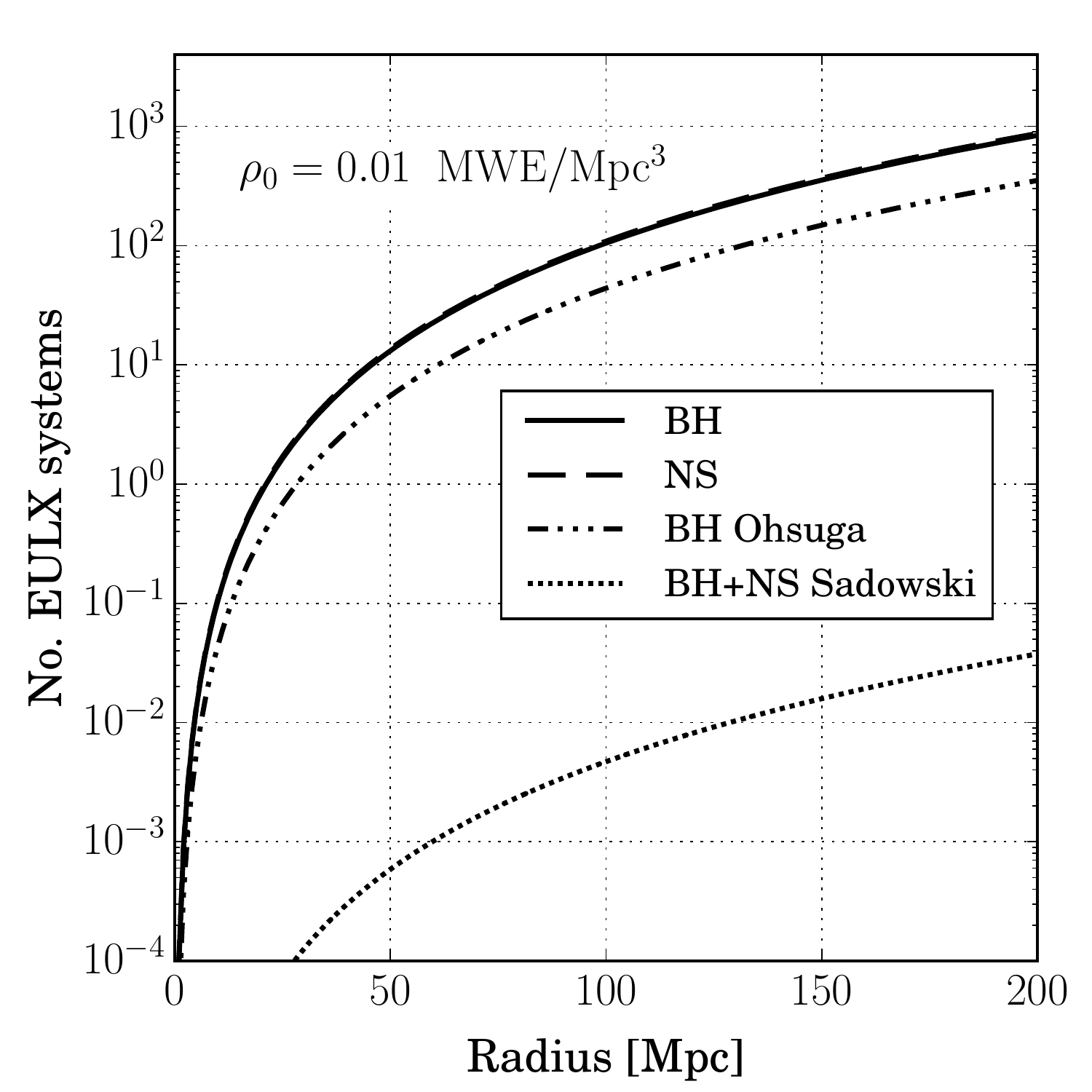}
    \caption{
        Upper limits on the number of EULXs within the sphere of radius $R$
    (solid for BH systems, dashed for NS systems) obtained under the assumption
    that a compact object can accrete at an arbitrary high rate (our reference
    model, Sec. \ref{sec:reference}). Note that had we imposed the classical 
    Eddington limit in our reference model, the predicted number of extreme ULXs 
    would be zero.
    The dot-dashed and dotted lines
    correspond to the limited accretion scenarios, with \lx\ calculated from
    the following models of \citetalias{Ohsuga07} and \citetalias{Sadowski15},
    respectively. 
    }
    \label{fig:number}
\end{figure}

Figure~\ref{fig:number} shows our upper limits for the number of EULXs
with BH and NS accretors. Note that in our reference model we obtain
virtually as many potential EULXs with a NS as with a BH accretor. This
estimate accounts for the specific lifetime of each binary in a potential EULX
phase. We assumed that the typical density of Milky Way Equivalent galaxies
(MWEG) in the local universe is $\rho_{\rm MWEG}=0.01$ Mpc$^{-3}$.  Currently,
the observations place the only confirmed EULX (HLX-1) at the distance of
$95\mpc$ \citep[][but see \citet{Lasota15}]{Wiersema10}. Our predicted upper
limit on the number of EULXs at this distance is $90$ and $93$ binaries with BH
and NS accretors, respectively. Both estimates significantly exceed the
observed number of sources.

Had we imposed the classical Eddington limit on the mass accretion rate in our
reference model, the number of EULXs would be zero for both types of accretors. 

\begin{figure}
    \centering
    \includegraphics[scale=0.5]{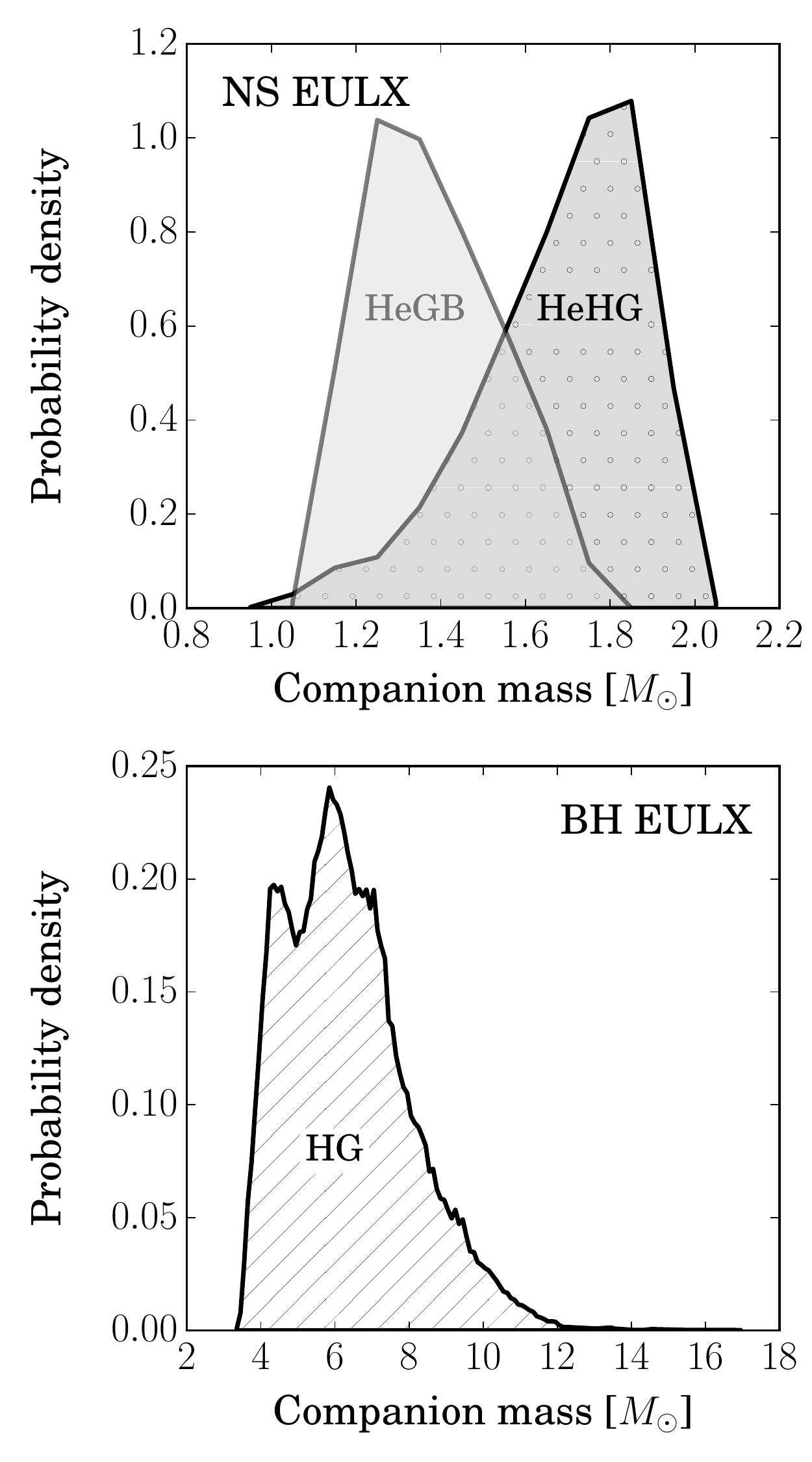}
    \caption{
        Mass distribution of the most common companions in NS (top) and BH
    (bottom) systems during the EULX phase in our reference model (Sec.
    \ref{sec:reference}). In the BH systems nearly all companions are on the
    Hertzsprung Gap (HG) with masses in the $4$--$10$\msun\ range, while the NS
    companions typically enter the EULX phase as Helium Hertzsprung Gap (HeHG)
    stars and evolve to Helium Giant Branch stars (HeGB).}
    \label{fig:mass}
\end{figure}

Table~\ref{tab:routes} introduces typical companion stars of our EULXs.
Majority of the NS EULXs (92\%) are found in binaries with Helium star donors:
either a $\sim$1.2\msun\ Helium Hertzsprung gap star (HeHG) or $\sim$1.8\msun\
Helium Giant Branch (HeGB) star.  The BH EULXs predominantly contain a
hydrogen-rich stars that have just evolved beyond main sequence.  Typically,
these are $\sim$6\msun\ Hertzsprung gap (HG) stars (92.4\%). The mass
distributions of these most common EULX companions are presented in
Fig.~\ref{fig:mass}. Table~\ref{tab:oldroutes} contains typical evolutionary
routes that lead to the formation of EULXs.

\begin{deluxetable*}{cccccccc}
    \tablewidth{\textwidth}
    \tablecaption{Compact accretors and their typical companions}
    \tablehead{Accretor type\tablenotemark{a}& \multicolumn{6}{c}{Companion type\tablenotemark{b}}}
    \startdata
    NS EULX & RG & EAGB & {\bf \new{HeHG}} & {\bf HeGB} & CO WD & ONe WD & \\
    $2.6\times10^{-3}$/MWEG & $6.35\%$ & $6.4\times10^{-3}\%$ & {\bf 48\%} & {\bf 44\%} & $0.5\%$ & $0.8\%$ & \\
& & & & & & & \\
    BH EULX & MS & {\bf HG} & RG & CHeB & EAGB & HeHG & HeGB \\
    $2.5\times10^{-3}$/MWEG & $3.4\%$ & {\bf 92.4\%} & $1\%$ & $0.3\%$ & $0.5\%$ & $0.5\%$ & $0.3\%$\\
    \enddata
    \label{tab:routes}
    \tablenotetext{a}{Type of accretor and the number of EULXs in the Milky Way
    Equivalent galaxy.}
    \tablenotetext{b}{Percentage of EULX systems with the same type of 
    accretor, MS~-~Main Sequence, HG~-~Hertzsprung Gap, RG~-~Red
    Giant, CHeB~-~Core Helium Burning, EAGB~-~Early Asymptotic Giant
    Branch, \new{HeHG}~-~Helium Hertzsprung gap, HeGB~-~Helium Giant Branch, CO
    WD~-~Carbon-Oxygen White Dwarf, ONe WD~-~Oxygen-Neon White Dwarf. }
\end{deluxetable*}

\begin{deluxetable*}{lccl}
    \tablewidth{\textwidth}
    \tablecaption{Typical EULX evolutionary routes}
        \tablehead{Accretor/route & \%\tablenotemark{a} & Number/MWEG\tablenotemark{b} &
        Evolutionary route\tablenotemark{c}}
    \startdata
    NS/1 & 50\% & $1.3\times10^{-3}$ & MT1(2/3-1) MT1(8/9-1) SN1 CE2(13-3/4;13-7) MT2(13-8/9)\\
    NS/2 & 40\% & $1.0\times10^{-3}$ & MT1(2/3-1) SN1 CE2(13-3/4;13-7) MT2(13-8/9)\\
    NS/3 & 10\% & $2.6\times10^{-4}$ & Other\\
    & & & \\
    BH/1 & 97\% & $2.5\times10^{-3}$ &CE1(4-1;7-1) SN1 MT2(14-1/2/3/4/5)\\
    BH/2 & 3\% & $6.8\times10^{-5}$ & Other\\
    \enddata
    \label{tab:oldroutes}
    \tablenotetext{a}{Percentage of the systems with the same type of 
    accretor.}
    \tablenotetext{b}{Number of systems expected to be observed per Milky-Way
    Equivalent galaxy.}
    \tablenotetext{c}{Symbolic designation of the evolutionary routes; MT1/MT2
    ~-~ mass transfer from the primary/secondary, SN1 ~-~ supernova explosion.
    CE1/CE2 ~-~ common envelope phase started by the primary/secondary. Numbers
        inside the parentheses specify the evolutionary phases of the stars:
1~-~MS; 2~-~HG; 3~-~RG; 4~-~CHeB; 5~-~EAGB; 7~-~HeMS; 8~-~HeHG; 9~-~HeGB;
13~-~NS; 14~-~BH (see \citet{Belczynski08} for details).}
\end{deluxetable*}

As a result of implementing the slim disk model of \citetalias{Ohsuga07}, the
X-ray luminosities of our systems drop by a factor of at least $2.8$.
Nevertheless, in this framework it is still possible to significantly exceed
the classical Eddington limit. The estimated number of the BH EULX systems
decreases to $1$ in $105$ billion binaries, and the number of BH EULXs
within $100\mpc$ is $37$ (Fig.~\ref{fig:number}).

After employing the model of \citetalias{Sadowski15} the upper limit on a
number of expected EULXs drops to 1 per $2\times10^{15}$ binaries.  The expected
upper limit on the number of EULX systems in the distance range of HLX-1 is
$0.004$.  This is several orders of magnitude lower than in the two other
models we tested.

In all cases, the EULX luminosities are achieved during the thermal timescale
RLOF. The thermal RLOF mass transfer rate is calculated following \citet{Kalogera96} as

\begin{equation}
    \mdot_{\rm RLOF,th} = {M_{\rm don} \over  \tau_{th}} = {1 \over 3} \times
    10^{-7} {R_{\rm don} L_{\rm don} \over M_{\rm don}} \msy,
\end{equation} 
where donor mass, radius and luminosity are expressed in solar units. For
example an EULX with a BH can start with a $\sim 10\msun$ Hertzsprung gap
donor, that has $\sim 20\rsun$ radius and luminosity of $\sim2\times10^4\lsun$.
The donors in NS systems are evolved (post core He-burning) low-mass Helium
stars that on the onset of RLOF have mass $\sim2\msun$, radius $\sim 30\rsun$
and luminosity $\sim 2\times 10^4\lsun$.  These parameters allow for very high
RLOF mass transfer rates ($\mdot_{\rm RLOF,th} \sim 10^{-2}$--$10^{-3}\msy$.
If this mass transfer is efficiently converted to X-ray luminosity
($\eta=\eta_0$), it will correspond to $L_{\rm x} \gtrsim
10^{42}$--$10^{43}\ergs$) alas on a very short timescale ($\tau_{\rm th} \sim
10^2$--$10^4$\,yr).

Typical evolutionary routes that lead to the formation of a potential BH and NS
EULX when $\eta=\eta_0$ are detailed in
\s\ref{sec:bh}--\ref{sec:ns}.  The corresponding time behaviour of the mass
accretion rate and maximum X-ray luminosity are shown in Figs.~\ref{fig:bh}
and~\ref{fig:ns}.

\subsection{Black hole EULX} \label{sec:bh}

A typical BH EULX system evolves along the evolutionary route BH/1
presented in Table~\ref{tab:oldroutes}. Its evolution begins with a $33 \msun$
primary and $11 \msun$ secondary on an orbit with separation $\sim5500 \rsun$
and eccentricity $e=0.56$.  In $5.5\myr$ the primary starts crossing the
Hertzsprung gap (HG). At that point the orbit has expanded to $5900\rsun$ due
to wind mass loss from the primary (now $30\msun$).  After about $10,000\yr$
and significant radial expansion ($1300\rsun$), the primary begins core helium
burning (CHeB). As the primary approaches its Roche lobe tidal interactions
circularise the orbit. After some additional expansion (radius $1700\rsun$) and
mass loss the primary ($18\msun$) initiates a common envelope (CE) phase.
Following the envelope ejection the orbit contracts to $40\rsun$ and the
primary becomes a Wolf-Rayet (WR) star with the mass of $11\msun$. After
$6.2\myr$ of evolution the primary undergoes a core collapse and forms a
$7.2\msun$ BH.  At this time the orbit is rather compact ($a=47\rsun$) and
almost circular ($e=0.04$). 

\begin{figure}
    \centering
    \includegraphics[scale=0.5]{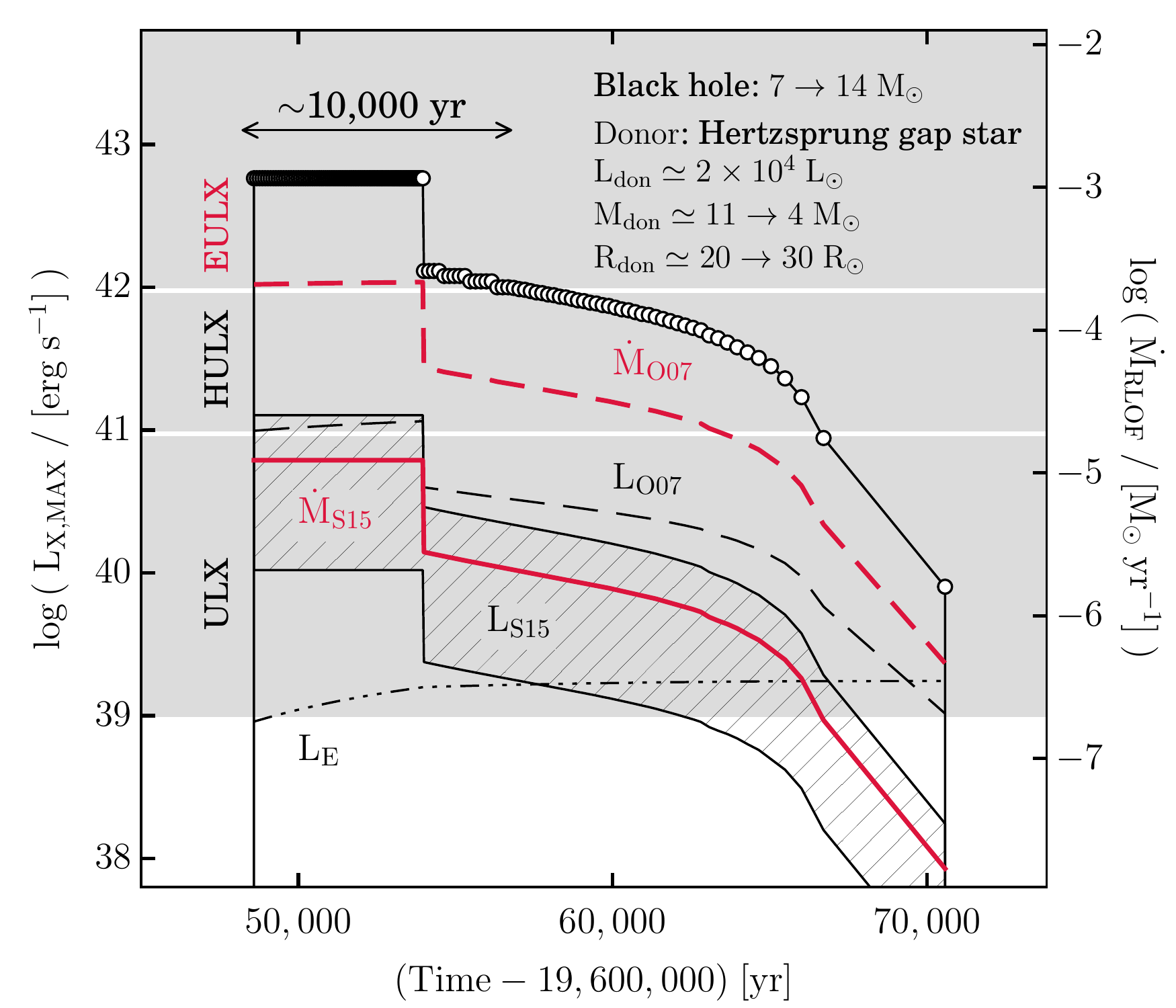}
    \caption{
    Typical evolution through Roche lobe overflow for a potential BH EULX
    binary in our reference model (open circles, $\eta=\eta_0$, see
    \s\ref{sec:reference}).  A $11\msun$Hertzsprung gap star transfers its
    H-rich envelope to a $7\msun$ black hole. The mass transfer during the EULX
    phase is driven on a thermal timescale of the donor ($\sim 10,000$yr) at a
    very high rate $\sim 10^{-3}\msy$ (see \s\ref{sec:bh}). The thick/red solid
    line and the hatched area represent, respectively, the mass accretion rate
    and the range of viewing angle dependent luminosities
    ($\theta=0^\circ$--$30^\circ$) obtained for our \mrlof\ time evolution with
    the model of \citetalias{Sadowski15} ($\facc=0.01$, $\fplus=8$). Similarly,
    the thick/red dashed line and thin/black dashed line show the mass
    accretion rate and luminosity derived with the model of
    \citetalias{Ohsuga07} ($\facc\simeq 0.2$--$0.3$, $\fplus\approx66$). The
    classical Eddington limit is marked with dot-dashed line.} \label{fig:bh}
\end{figure}

After the next $13\,\myr$ the secondary enters the HG with a mass of $11\msun$
and radius $10\rsun$, and expands filling its Roche lobe ($R_{\rm
lobe}=19\rsun$).  The mass transfer begins at orbital separation of $46\rsun$.
The luminosity of the donor is $21,000\lsun$.  Initially, for a short period of
time ($6,000\yr$) the mass transfer proceeds on the donor thermal timescale
with the RLOF mass transfer rate of $\mdot_{\rm RLOF}=1.2\times10^{-3}\msy$
(corresponding to $L_{\rm x} = 5.8 \times 10^{42} \ergs$). We allow the
entire transferred material to accrete onto the BH. After mass ratio reversal,
the orbit begins to expand in response to the mass transfer and the RLOF slows
down.  However, for the next $2,000\yr$ the donor RLOF rate stays above
$2.2\times10^{-4}\msy$ ($L_{\rm x} = 10^{42} \ergs$) and the system is
still classified as a potential EULX. The evolution of the mass transfer rate
throughout the RLOF phase is shown in Fig.~\ref{fig:bh}. 

The RLOF terminates when the secondary has transferred most of its H-rich
envelope to the BH (with the final mass $14\msun$). The reminder of the
secondary envelope is lost in a stellar wind and the secondary becomes a naked
helium star with the mass of $2.5\msun$. At this point the binary separation is
$230\rsun$. The low mass helium secondary ends its evolution in the Type Ib/c
supernova explosion forming a low mass NS ($\sim 1.1\msun$). The natal kick (if
significant) is very likely to disrupt the system.

\subsection{Neutron star EULX}\label{sec:ns}

A typical system (NS/1 channel in Table~\ref{tab:oldroutes}) begins as a
$10\msun$ primary and a $5.6\msun$ secondary on a $700\rsun$ orbit with an
eccentricity $e=0.73$. In $24\myr$ the primary enters the HG.  After
$50,000\yr$ of expansion on the HG the primary overfills its Roche lobe at the
radius of $85\rsun$. The orbit circularises and becomes $190\rsun$. We assume a
non-conservative mass transfer in such a case, and allow $50\%$ of transferred
material to accumulate on the main sequence secondary. At the end of this RLOF
episode the primary becomes a low-mass helium star ($2.2\msun$) and the
secondary becomes a rejuvenated main sequence star ($9.5\msun$). The separation
has expanded to $640\rsun$. The low mass helium star evolves and expands after
its core He burning is completed.  When the radius of the primary ($200\rsun$)
exceeds its Roche lobe, the second phase of a non-conservative mass transfer is
initiated. At this point the separation is $720\rsun$ and the masses are
$2.06\msun$ and $9.45$ for the primary and secondary, respectively. After a
short phase of a RLOF ($4,000\yr$) and after $28.6\myr$ since the zero age main
sequence (ZAMS), the primary explodes in electron capture supernova and forms a
low mass NS ($1.26\msun$).  We assume zero natal kick in this case and the
system survives the explosion. 

\begin{figure}
    \centering
    \includegraphics[scale=0.5]{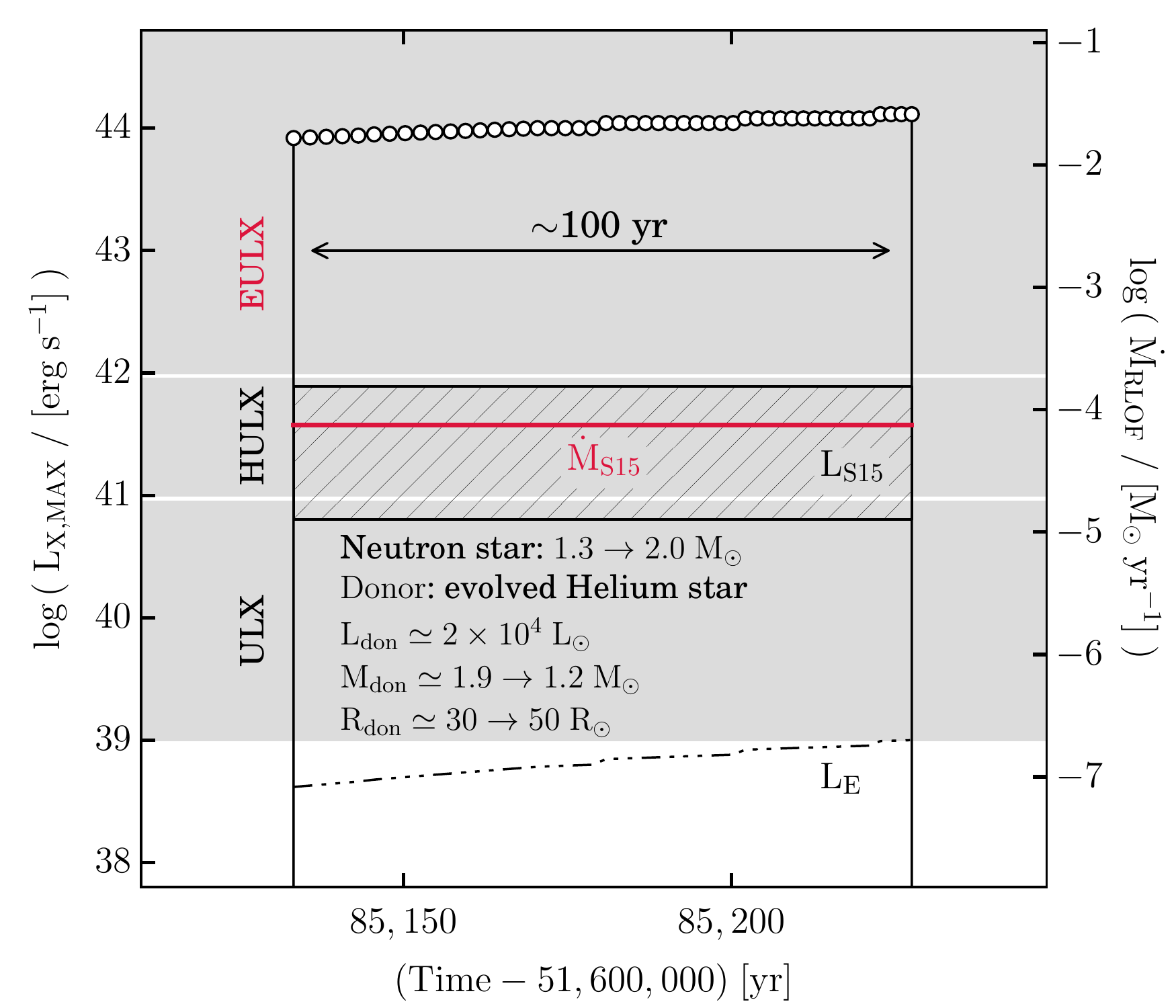}
    \caption{
    Typical evolution through Roche lobe overflow for a potential NS EULX
    binary system in our reference model (open circles, $\eta=\eta_0$, see
    \s\ref{sec:model_ohsuga}). A $1.9\msun$ evolved helium star transfers its
    He-rich envelope to a neutron star. The mass transfer is driven on a very
    short thermal timescale ($\sim100\yr$) at a rate $\sim10^{-2}\msy$ (see
    \s\ref{sec:ns}). The thick/red solid line and the hatched area represent,
    respectively, the mass accretion rate and range of viewing angle dependent
    luminosities obtained for our \mrlof\ evolution with the model of
    \citetalias{Sadowski15} ($\facc=0.01$, $\fplus\approx8$). Note that the
    model of \citetalias{Sadowski15} considers the case of BH accretion,
    however, in the case of a NS accretor, we assumed the same prescription for
    \lx.}
    \label{fig:ns}
\end{figure}

After additional $20\myr$ the $9.3\msun$ secondary leaves the main sequence and
evolves through the HG to become a red giant. Expansion on the red giant branch
leads to a RLOF that due to a large mass ratio and a deep convective envelope
of the secondary turns into a CE phase. At the onset of the CE, the separation
is $600\rsun$ and the secondary radius is $330\rsun$. We assume energy balance
for the CE \citep{Webbink84}, and obtain the post-CE separation of $50\rsun$.
The secondary has lost its entire H-rich envelope and becomes a low-mass helium
star ($2\msun$). After core helium burning, the low mass secondary begins to
expand and finally overfills its Roche lobe at the radius of $27\rsun$ (the
corresponding binary separation is $55\rsun$). At this point the evolved helium
star crosses the Helium Hertzsprung Gap and drives a high rate mass transfer
($\mdot_{\rm RLOF}=7.8\times10^{3}\msy$; $\lx=8\times10^{43} \ergs$) onto its
NS companion.  Since in our first approximation we have assumed a fully
efficient accretion with no limit, the NS mass quickly increases to $2\msun$.
The RLOF phase is very short ($100\yr$), and it drives an extremely high mass
transfer rate allowing this system to become a potential EULX with a NS
accretor (see Fig.~\ref{fig:ns}). Finally, the secondary is depleted of its
He-rich envelope and forms a naked CO core that cools off to become a CO WD.
After $51.6\myr$ since the ZAMS we note the formation of a wide NS-WD binary
(separation of $60\rsun$), with the gravitational merger time exceeding the
Hubble time ($t_{\rm merger}=2.8\times10^{14}\yr$).

\section{Discussion}\label{sec:discussion}

By allowing that all matter transferred through the RLOF is accreted (\macc
= \mrlof) and converted efficiently into X-rays ($\lx=\eta \macc c^2$, where
$\eta=\eta_0$ as in the standard disk), we are able to form a large number of
potential BH {\it and} NS EULX systems. The EULX phases that we obtain are
short ($\sim10,000 \yr$ and $\sim100\yr$ for BH and NS systems, respectively).
Nevertheless, we observe them in $0.1\%$ of all simulated binaries. Our
parameter space covers all progenitors of X-ray binaries, so every $1$ in
$1,000$ XRBs should become an EULX during its evolution. Luminosities that we
find in our simulations exceed $10^{42}\ergs$. Such high luminosities are found
both for BH and NS accretors.  Particularly, the presence of the potential
EULXs with NS accretors in our results seems to agree with the recent discovery
of the NS ULX system M82 X-2 \citep{Bachetti14}.

Evolution of ULXs with NS accretors was also the topic of a recent work by
\citet{Fragos15}. They used the BSE code for evolution of binaries and the MESA
code to calculate precisely the mass transfer phases. They found that NS ULX
systems should exist in 13\% of M82-like galaxies. They also found donors to be
H-rich stars with masses in the $3$--$8\msun$ range and $1$--$3$ day orbital
periods.  The orbits in our NS EULX systems are wider ($\sim30$ days periods)
and the companion stars are lighter ($1$--$2\msun$) and of a different type
(evolved helium stars) than those reported in \citet{Fragos15}.  They
obtained mass transfer rates up to $\mrlof\approx10^{-2}\msy$ (see their
Fig.~4), so they have reached mass transfer rates approximately as high as in
our study.  However, it needs to be noted that they have studied a much broader
population of ULXs, while we have focused only on the brightest ones. 

A problem of the maximum X-ray luminosity available from a binary system
was considered also by \citet{Podsiadlowski03}.  However, their study was
confined to only 11 evolutionary routes, only BH binary systems, and quite
limited parameter space (systems composed initially of a primary with mass in
the $25$ -- $45\msun$ range and a $2$ -- $17\msun$ secondary mass range).
\citet{Rappaport05} extended the simulations of \citet{Podsiadlowski03} and
were able to obtained ULXs with X-ray luminosities of $3\times10^{42}\ergs$
(see their Fig. 11) for a $5\msun$ BH and $9\msun$ donor and the case B mass
transfer (HG donor).  This is consistent with our typical BH EULX system.
Their secondaries also fill their Roche lobe due to expansion of the envelope,
and the ULX phase lasts at most a few \myr\ (about $10,000\yr$ as EULX).
However, their grid of parameters is far more sparse than ours, as they
simulated only 52 specific evolutionary cases in comparison to our $10^9$
evolutionary routes. 

\citet{Podsiadlowski03} and \citet{Rappaport05} used detailed evolutionary
code to obtained their results, while we have used simplified evolutionary
formulae to cover larger parameter space. Both approaches have their
advantages. With our approach we could not only confirm, but also extend the
previously published results, both in terms of the accretor type (NS accretors
possible) and in the range of potentially available mass transfer rates onto
compact accretors in close binary systems. 

Reference model scenario leads to a significant overestimate of
the number of the potential EULX systems in the local Universe.  We find close
to $\sim100$ systems with a NS, and a similar number of systems with a BH
within the $100\mpc$ radius. Note that these numbers should be considered as
upper limits.  However, to date observations have revealed only one system with
$\lx > 10^{42}$\ergs, HLX-1 \citep{Farrell11}.

Observational data indicates that HLX-1 is a transient system with
recurrent outbursts on timescales of about 1 year \citep[e.g.,][]{Godet09}.  Thermal-viscous
instability mechanism was proposed to drive outburst in XRBs. However, this
mechanism is not operational for mass transfer rates exceeding $(a\, few)\times
10^{-5} \msy$  because the disk becomes too hot and thus constantly ionised
\citep{Lasota15}. If the periodicity observed in HLX-1 is connected to
thermal-viscous instability, then the mass transfer rate needs to be lower
than the above threshold, and high X-ray luminosity is achieved by effective
beaming of radiation. 

Since our reference model provides only a very crude estimate of X-ray
luminosity, we proceeded with investigating the state-of-the-art global
accretion models of the super-Eddington accretion regime. \citetalias{Ohsuga07}
performed 2-D radiation hydrodynamic simulations of supercritical accretion
disks around BHs with the $\alpha P$ viscosity, while
\citetalias{Sadowski15} performed simulations of the magnetised accretion disks
in general relativity using radiation MHD code {\tt KORAL}. The former
simulations were fed by inflowing stream of gas circularising near $R_{\rm
cir}=100R_{\rm G}$, while the latter were initialised as equilibrium torii
threaded by seed magnetic field. Both models agree that there is significant
mass loss in the accretion flow, driven either by radiation pressure itself, or
by radiation pressure and the centrifugal force. The work by
\citetalias{Ohsuga07} provides a dependence of the mass accretion rate and
luminosity on the mass input rate (here $\dot{M}_{\rm RLOF}$).  However, this
result depends strongly on the assumed location of the circularization radius,
i.e., it implicitly assumes that no gas is lost outside $R\approx 100 R_{\rm
G}$. In real RLOF systems, the outer edge of the disk is expected to be located
much farther (up to about $2/3$ Roche lobe radius of the accretor). The other
approach \citepalias{Sadowski15} was not limited by the disk truncation inside
the simulation box, but rather by the computational time, which allowed flow to
reach the inflow/outflow equilibrium only to radius $R_{\rm eq}\approx 100
R_{\rm G}$. Inside this region, the gas flows out as wind down to $R_{\rm
wind}\approx 20 R_{\rm G}$, and shows roughly constant mass loss rate ${\rm
d}\dot M_{\rm wind}/{\rm d}R$. The total amount of gas lost in the system will
therefore depend on the location of the outer disk edge, or rather the outer
edge of the wind emitting region, through Eq.~\ref{eq:mdot_as}. Because of poor
understanding of the dynamics in the outer region of accretion disks, we
assumed $R_{\rm wind} / R_{\rm out} = 0.01$.

Within all of the discussed here accretion models binary systems are 
able to breach the $10^{42}\ergs$ EULX limit. Although, in the case of 
\citetalias{Sadowski15} the
probability of forming a binary EULX is a few orders of magnitude smaller than
for the other cases.  Limitation of the mass accretion rate onto the BH in
models developed in \citetalias{Ohsuga07} and \citetalias{Sadowski15} results
not only in lower X-ray luminosity, but also in different orbit evolution as
compared to our typical EULX case presented in Fig.~\ref{fig:bh}. Both effects
(lower X-ray luminosity and different orbit evolution) decrease the predicted
number of EULXs in the local Universe for \citetalias{Ohsuga07} and
\citetalias{Sadowski15} models.

If disk winds are effective, matter which cannot be accreted is ejected
from the system and takes away angular momentum. As a result the binary
separation decreases and this prevents longer phases of high mass transfer. For
the case of the \citetalias{Ohsuga07} outflow model, the total time spent in
the EULX regime is about $3,000\yr$, as opposed to $10,000\yr$  in our
reference scenario. As a result, we estimated upper limit on the number of the
potential BH EULXs drops in the \citetalias{Ohsuga07} model from $\sim 100$ to
$37$ within the $100\mpc$ radius. In Fig.~\ref{fig:bh} we show how the
$\dot{M}_{\rm RLOF}$ of our typical BH EULX (the reference model) translates
into $\lx$ and $\dot{M}_{\rm acc}$ derived from the model of
\citetalias{Ohsuga07}.

When we limit the $\dot{M}_{\rm acc}$ according to Eq.~\ref{eq:mdot_as}
\citepalias{Sadowski15}, we find considerably fewer EULXs within the $100\mpc$
radius than in the case of our reference model (factor of $\sim 5\times10^{4}$
fewer). For accretion limited case employing the model of
\citetalias{Ohsuga07} we obtain only a few times fewer number of EULXs (factor
of $\sim 3$ fewer).  Particularly, in the case of our typical BH EULX system
(Fig.~\ref{fig:bh}) we find 2 orders of magnitude smaller luminosity in the
beam than in the reference model.  In Fig.~\ref{fig:bh} we show the
$\dot{M}_{\rm acc}$ and a range of $L_{\rm x}$ computed for $\theta=0^\circ$ --
$30^\circ$ \citep[Eqs.~\ref{eq:mdot_as} and \ref{eq:l_as},
respectively.][]{Sadowski15} corresponding to the mass transfer history
$\dot{M}_{\rm RLOF}$ of our typical BH EULX (the reference model). 

Both \citetalias{Ohsuga07} and \citetalias{Sadowski15} considered accretion
disks around BHs.  The case of a NS accretion is far more complicated. A strong
magnetic fields associated with NSs may disrupt the disk far away from the
accretor.  Even if the magnetic field is weak, the emission will be weaker as
the accretor mass is lower. On the other hand, the NS accretion efficiency
($\eta$) is higher than that of BHs as the matter and photons do not fall under
the event horizon. To date, no comprehensive simulations of such a case have not been
performed.  \citet{Ohsuga07b} performed 2D simulations for a very limited range
of initial parameters.  They obtained supercritical accretion rates for NS
accretors and provided information that the mass accretion rate is 20-30\% of
that on to the BH for the same mass-input rate.  Thus, we have assumed NS
Eddington limited accretion case in the Oshuga et al. models, but we have
allowed super-Eddington accretion for S\k{a}dowski et al. models. 

We note that had we impose the logarithmic scaling of \lx\ with $\mdot_{\rm
acc}$ \citep[e.g.,][]{Poutanen07}, we would have not obtained any systems with
luminosities in the EULX regime. There exist factors that can improve this
situation like beaming of the radiation, however, spherical nebulae observed
around some of ULXs \citep{Pakull02,Russel11,Moon11} may suggest dispersion of
the outflow energy and nearly isotropic emission. The relation between
$\mdot_{\rm acc}$ and $\lx$ has not been derived from first principles, and
advanced numerical models do not seem to confirm this relation \citep[e.g.,][]{Ohsuga07,Sadowski15}.

\citet{Gladstone13} and \citet{Heida14} presented programs to search for the
companion stars of the ULX systems in the optical and infrared bands,
respectively. The former group investigated close ($\sim5\mpc$) ULXs and
discovered $13\pm5$ optical counterparts among $33$ ULXs. The masses of the
companions have large uncertainties and generally only the upper limit is
provided in the  mass range $5.7$--$16.1$\msun. In a few cases the lower limit
is also present in the mass range $8.3$--$14.7$\msun. These constrains are in
agreement with our results.  \citet{Heida14} investigated 62 close ULXs and
discovered 17 potential counterpart candidates.  Based on the absolute
magnitudes most of them (11) are Red Supergiants. According to our results the
most common companions of the ULX systems able to reach the EULX regime are
$6\msun$ Hertzsprung gap stars with BH accretors and $1$ -- $2\msun$ low mass
Helium stars for NS accretors.  However, it is possible that the Red
Supergiants that we found to be in minority among the EULX companions, are more
typical in the case of standard ULX systems. Detailed investigation of the
properties of the entire ULX population will be presented in our forthcoming
paper.

\section{Conclusions}

We conducted a proof-of-concept study to investigate if a binary system can
form an ULX with the extreme mass transfer rate potentially able to lead to the
X-ray luminosities in the $>10^{42}\ergs$ range. This is at least hundred times
more than what is expected for the Eddington limited stellar-origin $100\msun$ BH
\citep{Belczynski10, Belczynski14}. Observations of HLX-1 in ESO 243-49 with
X-ray luminosity of $1.1\times10^{42}\ergs$ encouraged us to look into the
problem. 

We find several evolutionary channels that lead to phases of a very high mass
transfer rate in close RLOF binaries. These evolutionary phases can be extremely 
short, but it appears that many binaries experience such phases. 
The mass transfer rate may so high that (if Eddington limit, or any similar limit 
is breached) the X-ray luminosity may reach well above $10^{42}\ergs$. 
We have adopted two physical accretion disk models in X-ray binaries, and we
have shown that for each model (that takes into account mass loss from
accretion flow and photon trapping) EULX sources are formed within typical
stellar populations. We note that increasing the geometrical beaming of
radiation, many more binaries than considered in our study could possibly
become ULX or EULX. 

It is found that about half of this potential extreme ULX systems host not
BH, but NS accretors. This contradict the present consensus in the community, as
the brightest ULX systems are commonly suspected to host IMBH primaries. Or at
the minimum with stellar-origin BHs.

\acknowledgements We would like to thank volunteers whose participation in the
universe@home test project\footnote{http://universeathometest.info} made it
possible to acquire the results in such a short time. This study was partially
supported by the Polish NCN grant N203 404939, Polish FNP professorial subsidy
"Master2013" and by Polish NCN grant SONATA BIS 2 (DEC-2012/07/E/ST9/01360).
MS was partially supported by the Polish NCN grant No. 2011/03/B/ST9/03459.  KB
also acknowledges NASA Grant Number NNX09AV06A and NSF Grant Number HRD 1242090
awarded to the Center for Gravitational Wave Astronomy, UTB.

\bibliographystyle{apj}
\bibliography{ms}

\end{document}